\documentstyle[12pt]{book}
\begin{document}
 
\sloppy
\raggedbottom
 
%% 25 Aug 99
 
\chapter*
{A Century of Controversy over the Foundations of Mathematics}
\markright
{A Century of Controversy over the Foundations of Mathematics}
\addcontentsline{toc}{chapter}
{A Century of Controversy over the Foundations of Mathematics}
 
\section*{G. J. Chaitin, \it chaitin@watson.ibm.com}
\section*{}
 
{\it
Lecture given Friday 30 April 1999 at UMass-Lowell.
The lecture was videotaped; this is an edited transcript.
}
 
\section*{Prof.\ Ray Gumb}
 
We're happy to have Gregory Chaitin from IBM's Thomas J. Watson
Research Lab to speak with us today.  He's a world-renowned figure,
and the developer as a teenager of the theory of algorithmic
information.  And his newest book which is accessible to
undergraduates, and I hope will be of great appeal to our
undergraduates in particular, is available on the Web and comes with
LISP programs to run with it.  It's kind of like a combination of
mathematics, computer science, and philosophy.  Greg---
 
\section*{Greg Chaitin}
 
Thanks a lot!  Okay, a great pleasure to be here!  [Applause] Thank
you very much!  I'm awfully sorry to be late!  You've got a beautiful
town here!  Those old brick buildings and the canals are really
breathtaking!  And thanks for being here for this talk!  It's such a
beautiful spring day---I think one has to be crazy to be indoors!
 
Okay, I'd like to talk about some crazy stuff.  The general idea is
that sometimes ideas are very powerful.  I'd like to talk about
theory, about the computer as a concept, a philosophical concept.
 
We all know that the computer is a very practical thing out there in
the real world!  It pays for a lot of our salaries, right?  But what
people don't remember as much is that really---I'm going to
exaggerate, but I'll say it---the computer was invented in order to
help to clarify a question about the foundations of mathematics, a
philosophical question about the foundations of mathematics.
 
Now that sounds absurd, but there's some truth in it.  There are
actually lots of threads that led to the computer, to computer
technology, which come from mathematical logic and from philosophical
questions about the limits and the power of mathematics.
 
The computer pioneer Turing was inspired by these questions.  Turing
was trying to settle a question of Hilbert's having to do with the
philosophy of mathematics, when he invented a thing called the Turing
machine, which is a mathematical model of a toy computer.  Turing did
this before there were any real computers, and then he went on to
actually build computers.  The first computers in England were built
by Turing.
 
And von Neumann, who was instrumental in encouraging the creation of
computers as a technology in the United States, (unfortunately as part
of the war effort, as part of the effort to build the atom bomb), he
knew Turing's work very well.  I learned of Turing by reading von
Neumann talking about the importance of Turing's work.
 
So what I said about the origin of the computer isn't a complete lie,
but it is a forgotten piece of intellectual history.  In fact, let me
start off with the final conclusion of this talk\ldots In a way, a lot
of this came from work of Hilbert.  Hilbert, who was a very well-known
German mathematician around the beginning of this century, had
proposed formalizing completely all of mathematics, all of
mathematical reasoning---deduction.  And this proposal of his is a
tremendous, glorious failure!
 
In a way, it's a spectacular failure.  Because it turned out that you
couldn't formalize mathematical reasoning.  That's a famous result of
G\"odel's that I'll tell you about, done in 1931.
 
But in another way, Hilbert was really right, because formalism has
been the biggest success of this century.  Not for reasoning, not for
deduction, but for programming, for calculating, for computing, that's
where formalism has been a tremendous success.  If you look at work by
logicians at the beginning of this century, they were talking about
formal languages for reasoning and deduction, for doing mathematics
and symbolic logic, but they also invented some early versions of
programming languages.  And {\bf these} are the formalisms that we all
live with and work with now all the time!  They're a tremendously
important technology.
 
So formalism for reasoning did not work.  Mathematicians don't reason
in formal languages.  But formalism for computing, programming
languages, are, in a way, what was right in the formalistic vision
that goes back to Hilbert at the beginning of this century, which was
intended to clarify epistemological, philosophical questions about
mathematics.
 
So I'm going to tell you this story, which has a very surprising
outcome.  I'm going to tell you this surprising piece of intellectual
history.
 
\section*{The Crisis in Set Theory}
 
So let me start roughly a hundred years ago, with Cantor\ldots
\[
   \mbox{Georg Cantor}
\]
 
The point is this.  Normally you think that pure mathematics is
static, unchanging, perfect, absolutely correct, absolute truth\ldots
Right?  Physics may be tentative, but math, things are certain there!
Well, it turns out that's not exactly the case.
 
In this century, in this past century there was a lot of controversy
over the foundations of mathematics, and how you should do math, and
what's right and what isn't right, and what's a valid proof.  Blood
was almost shed over this\ldots People had terrible fights and ended
up in insane asylums over this.  It was a fairly serious controversy.
This isn't well known, but I think it's an interesting piece of
intellectual history.
 
More people are aware of the controversy over relativity theory.
Einstein was very controversial at first.  And then of the controversy
over quantum mechanics\ldots These were the two revolutions in the
physics of this century.  But what's less well known is that there
were tremendous revolutions and controversies in pure mathematics too.
I'd like to tell you about this.  It really all starts in a way from
Cantor.
\[
   \mbox{Georg Cantor}
\]
What Cantor did was to invent a theory of infinite sets.
\[
   \mbox{Infinite Sets}
\]
He did it about a hundred years ago; it's really a little more than a
hundred years ago.  And it was a tremendously revolutionary theory, it
was {\bf extremely} adventurous.  Let me tell you why.
 
Cantor said, let's take 1, 2, 3, \ldots
\[
   1, 2, 3, \ldots
\]
We've all seen these numbers, right?!  And he said, well, let's add an
infinite number after this.
\[
   1, 2, 3, \ldots
   \omega
\]
He called it $\omega$, lowercase Greek omega.  And then he said, well,
why stop here?  Let's go on and keep extending the number series.
\[
   1, 2, 3, \ldots
   \omega, \omega+1, \omega+2, \ldots
\]
Omega plus one, omega plus two, then you go on for an infinite amount
of time.  And what do you put afterwards?  Well, two omega?
(Actually, it's omega times two for technical reasons.)
\[
   1, 2, 3, \ldots
   \omega \ldots 2\omega
\]
Then two omega plus one, two omega plus two, two omega plus three, two
omega plus four\ldots
\[
   1, 2, 3, \ldots
   2\omega, 2\omega+1, 2\omega+2, 2\omega+3, 2\omega+4, \ldots
\]
Then you have what?  Three omega, four omega, five omega, six omega,
\ldots
\[
   1, 2, 3, \ldots
   3\omega \ldots 4\omega \ldots 5\omega \ldots 6\omega \ldots
\]
Well, what will come after all of these?  Omega squared!  Then you
keep going, omega squared plus one, omega squared plus six omega plus
eight\ldots Okay, you keep going for a long time, and the next
interesting thing after omega squared will be?  Omega cubed!  And then
you have omega to the fourth, omega to the fifth, and much later?
\[
   1, 2, 3, \ldots
   \omega \ldots \omega^2 \ldots \omega^3 \ldots \omega^4
   \ldots \omega^5
\]
Omega to the omega!
\[
   1, 2, 3, \ldots
   \omega \ldots \omega^2 \ldots \omega^{\omega}
\]
And then much later it's omega to the omega to the omega an infinite
number of times!
\[
   1, 2, 3, \ldots
   \omega \ldots \omega^2 \ldots \omega^{\omega} \ldots
   \omega^{\omega^{\omega^{\omega^{\ldots}}}}
\]
I think this is usually called epsilon nought.
\[
   \varepsilon_0 =
   \omega^{\omega^{\omega^{\omega^{\ldots}}}}
\]
It's a pretty mind-boggling number!  After this point things get a
little complicated\ldots
 
And this was just one little thing that Cantor did as a warm-up
exercise for his main stuff, which was measuring the size of infinite
sets!  It was spectacularly imaginative, and the reactions were
extreme.  Some people loved what Cantor was doing, and some people
thought that he should be put in an insane asylum!  In fact he had a
nervous breakdown as a result of those criticisms.  Cantor's work was
very influential, leading to point-set topology and other abstract
fields in the mathematics of the twentieth century.  But it was also
very controversial.  Some people said, it's theology, it's not real,
it's a fantasy world, it has nothing to do with serious math!  And
Cantor never got a good position and he spent his entire life at a
second-rate institution.
 
\section*{Bertrand Russell's Logical Paradoxes}
 
Then things got even worse, due mainly, I think, to Bertrand Russell,
one of my childhood heroes.
\[
   \mbox{Bertrand Russell}
\]
Bertrand Russell was a British philosopher who wrote beautiful essays,
very individualistic essays, and I think he got the Nobel prize in
literature for his wonderful essays.  Bertrand Russell started off as
a mathematician and then degenerated into a philosopher and finally
into a humanist; he went downhill rapidly!  [Laughter] Anyway,
Bertrand Russell discovered a whole bunch of disturbing paradoxes,
first in Cantor's theory, then in logic itself.  He found cases where
reasoning that seemed to be okay led to contradictions.
 
And I think that Bertrand Russell was tremendously influential in
spreading the idea that there was a serious crisis and that these
contradictions had to be resolved somehow.  The paradoxes that Russell
discovered attracted a great deal of attention, but strangely enough
only one of them ended up with Russell's name on it!  For example, one
of these paradoxes is called the Burali-Forti paradox, because when
Russell published it he stated in a footnote that it had been
suggested to him by reading a paper by Burali-Forti.  But if you look
at the paper by Burali-Forti, you don't see the paradox!
 
But I think that the realization that something was seriously wrong,
that something was rotten in the state of Denmark, that reasoning was
bankrupt and something had to be done about it pronto, is due
principally to Russell.  Alejandro Garciadiego, a Mexican historian of
math, has written a book which suggests that Bertrand Russell really
played a much bigger role in this than is usually realized: Russell
played a key role in formulating not only the Russell paradox, which
bears his name, but also the Burali-Forti paradox and the Berry
paradox, which don't.  Russell was instrumental in discovering them
and in realizing their significance.  He told everyone that they were
important, that they were not just childish word-play.
 
Anyway, the best known of these paradoxes is called the Russell
paradox nowadays.  You consider the set of all sets that are not
members of themselves.  And then you ask, ``Is this set a member of
itself or not?''  If it is a member of itself, then it shouldn't be,
and vice versa!  It's like the barber in a small, remote town who
shaves all the men in the town who don't shave themselves.  That seems
pretty reasonable, until you ask ``Does the barber shave himself?''
He shaves himself if and only if he doesn't shave himself, so he can't
apply that rule to himself!
 
Now you may say, ``Who cares about this barber!''  It was a silly rule
anyway, and there are always exceptions to the rule!  But when you're
dealing with a {\bf set}, with a mathematical concept, it's not so
easy to dismiss the problem.  Then it's not so easy to shrug when
reasoning that seems to be okay gets you into trouble!
 
By the way, the Russell paradox is a set-theoretic echo of an earlier
paradox, one that was known to the ancient Greeks and is called the
Epimenides paradox by some philosophers.  That's the paradox of the
liar: ``This statement is false!''  ``What I'm now saying is false,
it's a lie.''  Well, is it false?  If it's false, if something is
false, then it doesn't correspond with reality.  So if I'm saying this
statement is false, that means that it's not false---which means that
it must be true.  But if it's true, and I'm saying it's false, then it
must be false!  So whatever you do you're in trouble!
 
So you can't get a definite logical truth value, everything flip
flops, it's neither true nor false.  And you might dismiss this and
say that these are just meaningless word games, that it's not serious.
But Kurt G\"odel later built his work on these paradoxes, and he had a
very different opinion.
\[
   \mbox{Kurt G\"odel}
\]
He said that Bertrand Russell made the amazing discovery that our
logical intuitions, our mathematical intuitions, are
self-contradictory, they're inconsistent!  So G\"odel took Russell
very seriously, he didn't think that it was all a big joke.
 
Now I'd like to move on and tell you about David Hilbert's rescue plan
for dealing with the crisis provoked by Cantor's set theory and by
Russell's paradoxes.
\[
   \mbox{David Hilbert}
\]
 
\section*{David Hilbert to the Rescue with Formal Axiomatic Theories}
 
One of the reactions to the crisis provoked by Cantor's theory of infinite
sets, one of the reactions was, well, let's escape into {\bf
formalism}.  If we get into trouble with reasoning that seems okay,
then one solution is to use symbolic logic, to create an artificial
language where we're going to be very careful and say what the rules
of the game are, and make sure that we don't get the contradictions.
Right?  Because here's a piece of reasoning that looks okay but it
leads to a contradiction.  Well, we'd like to get rid of that.  But
natural language is ambiguous---you never know what a pronoun refers
to.  So let's create an artificial language and make things very, very
precise and make sure that we get rid of all the contradictions!  So
this was the notion of formalism.
\[
   \mbox{Formalism}
\]
 
Now I don't think that Hilbert actually intended that mathematicians
should work in such a perfect artificial language.  It would sort of
be like a programming language, but for reasoning, for doing
mathematics, for deduction, not for computing, that was Hilbert's
idea.  But he never expressed it that way, because there were no
programming languages back then.
 
So what are the ideas here?  First of all, Hilbert stressed the
importance of the axiomatic method.
\[
   \mbox{Axiomatic Method}
\]
The notion of doing mathematics that way goes back to the ancient
Greeks and particularly to Euclidean geometry, which is a beautifully
clear mathematical system.  But that's not enough; Hilbert was also
saying that we should use symbolic logic.
\[
   \mbox{Symbolic Logic}
\]
And symbolic logic also has a long history: Leibniz, Boole, Frege,
Peano\ldots These mathematicians wanted to make reasoning like
algebra.  Here's how Leibniz put it: He talked about avoiding
disputes---and he was probably thinking of political disputes and
religious disputes---by calculating who was right instead of arguing
about it!  Instead of fighting, you should be able to sit down at a
table and say, ``Gentleman, let us compute!''  What a beautiful
fantasy!\ldots
 
So the idea was that mathematical logic should be like arithmetic and
you should be able to just grind out a conclusion, no uncertainty, no
questions of interpretation.  By using an artificial math language
with a symbolic logic you should be able to achieve {\it perfect
rigor.}  You've heard the word ``rigor'', as in ``rigor mortis'',
used in mathematics?  [Laughter] It's not that rigor!  But the idea is
that an argument is either completely correct or else it's total
nonsense, with nothing in between.  And a proof that is formulated in
a formal axiomatic system should be absolutely clear, it should be
completely sharp!
 
In other words, Hilbert's idea was that we should be completely
precise about what the rules of the game are, and about the
definitions, the elementary concepts, and the grammar and the
language---all the rules of the game---so that we can all agree on how
mathematics should be done.  In practice it would be too much work to
use such a formal axiomatic system, but it would be philosophically
significant because it would settle once and for all the question of
whether a piece of mathematical reasoning is correct or incorrect.
 
Okay?  So Hilbert's idea seemed fairly straightforward.  He was just
following the axiomatic and the formal traditions in mathematics.
Formal as in formalism, as in using formulas, as in calculating!  He
wanted to go all the way, to the very end, and formalize all of
mathematics, but it seemed like a fairly reasonable plan.  Hilbert
wasn't a revolutionary, he was a conservative\ldots The amazing thing,
as I said before, was that it turned out that Hilbert's rescue plan
{\bf could not work}, that it couldn't be done, that it was {\bf
impossible} to make it work!
 
Hilbert was just following the whole mathematics tradition up to that
point: the axiomatic method, symbolic logic, formalism\ldots He wanted
to avoid the paradoxes by being absolutely precise, by creating a
completely formal axiomatic system, an artificial language, that
avoided the paradoxes, that made them impossible, that {\bf outlawed}
them!  And most mathematicians probably thought that Hilbert was
right, that {\it of course\/} you could do this---it's just the notion
that in mathematics things are absolutely clear, black or white, true
or false.
 
So Hilbert's idea was just an extreme, an exaggerated version of the
normal notion of what mathematics is all about: the idea that we can
decide and agree on the rules of the game, all of them, once and for
all.  The big surprise is that it turned out that this {\bf could not}
be done.  Hilbert turned out to be wrong, but wrong in a tremendously
fruitful way, because he had asked a very good question.  In fact, by
asking this question he actually created an entirely new field of
mathematics called {\bf meta}mathematics.
\[
   \mbox{Metamathematics}
\]
Metamathematics is mathematics turned inward, it's an introspective
field of math in which you study what mathematics can achieve or can't
achieve.
 
\section*{What is Metamathematics?}
 
That's my field---metamathematics!  In it you look at mathematics from
above, and you use mathematical reasoning to discuss what mathematical
reasoning can or cannot achieve.  The basic idea is this: Once you
entomb mathematics in an artificial language {\it \`a la\/} Hilbert,
once you set up a completely formal axiomatic system, then you can
forget that it has any meaning and just look at it as a game that you
play with marks on paper that enables you to deduce theorems from
axioms.  You can forget about the meaning of this game, the game of
mathematical reasoning, it's just combinatorial play with symbols!
There are certain rules, and you can study these rules and forget that
they have any meaning!
 
What things do you look at when you study a formal axiomatic system
from above, from the outside?  What kind of questions do you ask?
 
Well, one question you can ask is if you can prove that
``0 equals 1''?
\[
   0 = 1 \, ?
\]
Hopefully you can't, but how can you be sure?  It's hard to be sure!
 
And for any question $A$, for any affirmation $A$, you can ask if it's
possible to settle the matter by either proving $A$ or the opposite of
$A$, not $A$.
\[
   A \, ? \; \; \neg A \, ?
\]
That's called {\it completeness.}
\[
   \mbox{Completeness}
\]
A formal axiomatic system is complete if you can settle any question
$A$, either by proving it $(A)$, or by proving that it's false $(\neg
A)$.  That would be nice!  Another interesting question is if you can
prove an assertion $(A)$ and you can also prove the contrary assertion
$(\neg A)$.  That's called {\it inconsistency,} and if that happens
it's very bad!  {\it Consistency\/} is much better than {\it
inconsistency\/}!
\[
   \mbox{Consistency}
\]
 
So what Hilbert did was to have the remarkable idea of creating a new
field of mathematics whose subject would be mathematics itself.  But
you can't do this until you have a completely formal axiomatic system.
Because as long as any ``meaning'' is involved in mathematical
reasoning, it's all subjective.  Of course, the reason we do
mathematics is because it has meaning, right?  But if you want to be
able to study mathematics, the power of mathematics, using
mathematical methods, you have to ``desiccate'' it to ``crystallize
out'' the meaning and just be left with an artificial language with
completely precise rules, in fact, with one that has a {\bf mechanical
proof-checking algorithm}.
\[
    \mbox{Proof-Checking Algorithm}
\]
 
The key idea that Hilbert had was to envision this perfectly
desiccated or crystallized axiomatic system for all of mathematics, in
which the rules would be so precise that if someone had a proof there
would be a referee, there would be a mechanical procedure, which would
either say ``This proof obeys the rules'' or ``This proof is wrong;
it's breaking the rules''.  That's how you get the criterion for
mathematical truth to be completely objective and not to depend on
meaning or subjective understanding: by reducing it all to
calculation.  Somebody says ``This is a proof'', and instead of having
to submit it to a human referee who takes two years to decide if the
paper is correct, instead you just give it to a machine.  And the
machine eventually says ``This obeys the rules'' or ``On line 4
there's a misspelling'' or ``This thing on line 4 that supposedly
follows from line 3, actually doesn't''.  And that would be the end,
no appeal!
 
The idea was not that mathematics should actually be done this way.  I
think that that's calumny, that's a false accusation.  I don't think
that Hilbert really wanted to turn mathematicians into machines.  But
the idea was that if you could take mathematics and do it this way,
then you could use mathematics to study the power of mathematics.  And
{\bf that} is the important new thing that Hilbert came up with.
Hilbert wanted to do this in order to reaffirm the traditional view of
mathematics, in order to justify himself\ldots
 
He proposed having one set of axioms and this formal language, this
formal system, which would include all of mathematical reasoning, that
we could all agree on, and that would be {\bf perfect}!  We'd then
know all the rules of the game.  And he just wanted to use
metamathematics to show that this formal axiomatic system was
good---that it was consistent and that it was complete---in order to
convince people to accept it.  This would have settled once and for
all the philosophical questions ``When is a proof correct?''  and
``What is mathematical truth?''  Like this everyone could agree on
whether a mathematical proof is correct or not.  And in fact we used
to think that this was an objective thing.
 
In other words, Hilbert's just saying, if it's really objective, if
there's no subjective element, and a mathematical proof is either true
or false, well, then there should be certain rules for deciding that
and it shouldn't depend, if you fill in all the details, it shouldn't
depend on interpretation.  It's important to fill in all the
details---that's the idea of mathematical logic, to ``atomize''
mathematical reasoning into such tiny steps that {\bf nothing} is left
to the imagination, nothing is left out!  And if nothing is left out,
then a proof can be checked automatically, that was Hilbert's point,
that's really what symbolic logic is all about.
 
And Hilbert thought that he was actually going to be able to do this.
He was going to formalize all of mathematics, and we were all going to
agree that these were in fact the rules of the game.  Then there'd be
just one version of mathematical truth, not many variations.  We don't
want to have a German mathematics and a French mathematics and a
Swedish mathematics and an American mathematics, no, we want a {\bf
universal} mathematics, one universal criterion for mathematical
truth!  Then a paper that is done by a mathematician in one country
can be understood by a mathematician in another country.  Doesn't that
sound reasonable?!  So you can imagine just how very, very shocking it
was in 1931 when Kurt G\"odel showed that it wasn't at all reasonable,
that it could {\bf never} be done!
\[
    \mbox{1931 Kurt G\"odel}
\]
 
\section*{Kurt G\"odel Discovers Incompleteness}
 
G\"odel did this is Vienna, but he was from what I think is now called
the Czech republic, from the city of Br\"unn or Brno.  It was part of
the Austro-Hungarian empire then, but now it's a separate country.
And later he was at the Institute for Advanced Study in Princeton,
where I visited his grave a few weeks ago.  And the current owner of
G\"odel's house was nice enough to invite me in when he saw me
examining the house [laughter] instead of calling the police!  They
know they're in a house that some people are interested in for
historical reasons.
 
Okay, so what did Kurt G\"odel do?  Well, G\"odel sort of exploded
this whole view of what mathematics is all about.  He came up with a
famous incompleteness result, ``G\"odel's incompleteness theorem''.
\[
    \mbox{Incompleteness}
\]
And there's a lovely book explaining the way G\"odel originally did
it.  It's by Nagel and Newman, and it's called {\it G\"odel's
Proof.}  I read it when I was a child, and forty years later it's
still in print!
 
What is this amazing result of G\"odel's?  G\"odel's amazing discovery
is that Hilbert was {\bf wrong}, that it cannot be done, that there's
{\bf no way} to take all of mathematical truth and to agree on a set
of rules and to have a formal axiomatic system for all of mathematics
in which it is crystal clear whether something is correct or not!
 
More precisely, what G\"odel discovered was that if you just try to
deal with elementary arithmetic, with 0, 1, 2, 3, 4\ldots and with
addition and multiplication
\[
    + \; \times \; 0, 1, 2, 3, 4, \ldots
\]
---this is ``elementary number theory'' or ``arithmetic''---and you
just try to have a set of axioms for this---the usual axioms are
called Peano arithmetic---even this can't be done!  Any set of axioms
that tries to have {\bf the whole truth} and {\bf nothing but the
truth} about addition, multiplication, and 0, 1, 2, 3, 4, 5, 6, 7, 8,
9, 10\ldots will have to be incomplete.  More precisely, it'll either
be inconsistent or it'll be incomplete.  So if you assume that it only
tells the truth, then it won't tell the whole truth.  There's no way
to capture all the truth about addition, multiplication, and 0, 1, 2,
3, 4\ldots ! In particular, if you assume that the axioms don't allow
you to prove false theorems, then it'll be incomplete, there'll be
true theorems that you cannot prove from these axioms!
 
This is an absolutely devastating result, and all of traditional
mathematical philosophy ends up in a heap on the floor!  At the time
this was considered to be absolutely devastating.  However you may
notice that in 1931 there were also a few other problems to worry
about.  The situation in Europe was bad.  There was a major
depression, and a war was brewing.  I agree, not all problems are
mathematical!  There's more to life than epistemology!  But you begin
to wonder, well, if the traditional view of mathematics isn't correct,
then {\bf what is correct?} G\"odel's incompleteness theorem was very
surprising and a terrible shock.
 
How did G\"odel do it?  Well, G\"odel's proof is very clever.  It
almost looks crazy, it's very paradoxical.  G\"odel starts with the
paradox of the liar, ``I'm false!'', which is neither true nor false.
\[
    \mbox{``This statement is false!''}
\]
And what G\"odel does is to construct a statement that says of itself
``I'm unprovable!''
\[
    \mbox{``This statement is unprovable!''}
\]
Now if you can construct such a statement in elementary number theory,
in arithmetic, a mathematical statement---I don't know how you make a
mathematical statement say it's unprovable, you've got to be very
clever---but if you can do it, it's easy to see that you're in
trouble.  Just think about it a little bit.  It's easy to see that
you're in trouble.  Because if it's provable, it's false, right?  So
you're in trouble, you're proving false results.  And if it's
unprovable and it says that it's unprovable, then it's true, and
mathematics is incomplete.  So either way, you're in trouble!  Big
trouble!
 
And G\"odel's original proof is very, very clever and hard to
understand.  There are a lot of complicated technical details.  But if
you look at his original paper, it seems to me that there's a lot of
LISP programming in it, or at least something that looks a lot like
LISP programming.  Anyway, now we'd call it LISP programming.
G\"odel's proof involves defining a great many functions recursively,
and these are functions dealing with lists, which is precisely what
LISP is all about.  So even though there were no programming languages
in 1931, with the benefit of hindsight you can clearly see a
programming language in G\"odel's original paper.  And the programming
language I know that's closest to it is LISP, pure LISP, LISP without
side-effects, interestingly enough---that's the heart of LISP.
 
So this was a very, very shocking result, and people didn't really
know what to make of it.
 
Now the next major step forward comes only five years later, in 1936,
and it's by Alan Turing.
\[
    \mbox{1936 Alan Turing}
\]
 
\section*{Alan Turing Discovers Uncomputability}
 
Turing's approach to all these questions is completely different from
G\"odel's, and much deeper.  Because Turing brings it out of the
closet!  [Laughter] What he brings out of the closet is the computer!
The computer was implicit in G\"odel's paper, but this was really not
visible to any ordinary mortal, not at that time, only with hindsight.
And Turing really brings it out in the open.
 
Hilbert had said that there should be a ``mechanical procedure'' to
decide if a proof obeys the rules or not.  And Hilbert never clarified
what he meant by a mechanical procedure, it was all words.  But,
Turing said, what you really mean is a machine, and a machine of a
kind that we now call a Turing machine---but it wasn't called that in
Turing's original paper.  In fact, Turing's original paper contains a
programming language, just like G\"odel's paper does, what we would
now call a programming language.  But the two programming languages
are very different.  Turing's programming language isn't a high-level
language like LISP, it's more like a machine language.  In fact, it's
a horrible machine language, one that nobody would want to use today,
because it's too simple.
 
But Turing makes the point that even though Turing machines are very
simple, even though their machine language is rather primitive,
they're very flexible, very general-purpose machines.  In fact, he
claims, any computation that a human being can perform, should be {\bf
possible} to do using such a machine.  Turing's train of thought now
takes a very dramatic turn.  What, he asks, is {\bf impossible} for
such a machine?  What can't it do?  And he immediately finds a
question that no Turing machine can settle, a problem that no Turing
machine can solve.  That's {\it the halting problem,} the problem of
deciding in advance if a Turing machine or a computer program will
eventually halt.
\[
   \mbox{The Halting Problem}
\]
 
So the shocking thing about this 1936 paper is that first of all he
comes up with the notion of a general-purpose or universal computer,
with a machine that's flexible, that can do what any machine can do.
One calculating machine that can do {\bf any} calculation, which is,
we now say, a general-purpose computer.  And then he immediately shows
that there are limits to what such a machine can do.  And how does he
find something that cannot be done by any such machine?  Well, it's
very simple!  It's the question of whether a computer program will
eventually halt, with no time limit.
 
If you put a time limit, it's very easy.  If you want to know if a
program halts in a year, you just run it for a year, and either it
halted or doesn't.  What Turing showed is that you get in terrible
trouble if there's no time limit.  Now you may say, ``What good is a
computer program that takes more than a year, that takes more than a
thousand years?!  There's always a time limit!''  I agree, this is
pure math, this is not the real world.  You only get in trouble with
{\bf infinity}!  But Turing shows that if you put no time limit, then
you're in real difficulties.
 
So this is called {\it the halting problem.} And what Turing showed is
that there's no way to decide in advance if a program will eventually
halt.
\[
   \mbox{The Halting Problem}
\]
If it {\bf does} halt, by running it you can eventually discover that,
if you're just patient.  The problem is you don't know when to give
up.  And Turing was able to show with a very simple argument which is
just Cantor's diagonal argument---coming from Cantor's theory of
infinite sets, by the way---I don't have time to explain all
this---with a very simple argument Turing was able to show that this
problem
\[
   \mbox{The Halting Problem}
\]
cannot be solved.
 
No computer program can tell you in advance if another computer
program will eventually halt or not.  And the problem is the ones that
don't halt, that's really the problem.  The problem is knowing when to
give up.
 
So now the interesting thing about this is that Turing immediately
deduces as a corollary that if there's no way to decide in advance
{\bf by a calculation} if a program will halt or not, well then there
cannot be any way to {\bf deduce} it in advance using reasoning
either.  No formal axiomatic system can enable you to deduce in
advance whether a program will halt or not.
 
Because if you can use a formal axiomatic system to always deduce
whether a program will halt or not, well then, that will give you a
way to calculate in advance whether a program will halt or not.  You
simply run through all possible deductions---you can't do this in
practice---but in principle you can run through all possible proofs in
size order, checking which ones are correct, until either you find a
proof that the program will halt eventually or you find a proof that
it's never going to halt.
 
This is using the idea of a completely formal axiomatic system where
you don't need a mathematician---you just run through this calculation
on a computer---it's mechanical to check if a proof is correct or not.
So if there were a formal axiomatic system which always would enable
you to prove, to deduce, whether a program will halt or not, that
would give you a way to calculate in advance whether a program will
halt or not.  And that's impossible, because you get into a paradox
like ``This statement is false!''  You get a program that halts if and
only if it doesn't halt, that's basically the problem.  You use an
argument having the same flavor as the Russell paradox.
 
So Turing went more deeply into these questions than G\"odel.  As a
student I read G\"odel's proof, and I could follow it step by step: I
read it in Nagel and Newman's book, which is a lovely book.  It's a
marvelous book, it's so understandable!  It's still in print, and it
was published in 1958\ldots But I couldn't really feel that I was
coming to grips with G\"odel's proof, that I could really understand
it.  The whole thing seemed too delicate, it seemed too fragile, it
seemed too superficial\ldots And there's this business in the closet
about computing, that's there in G\"odel, but it's hidden, it's not in
the open, we're not really coming to terms with it.
 
Now Turing is really going, I think, much deeper into this whole
matter.  And he's showing, by the way, that it's not just one
particular axiomatic system, the one that G\"odel studied, that can't
work, but that {\bf no} formal axiomatic system can work.  But it's in
a slightly different context.  G\"odel was really looking at 0, 1, 2,
3, 4\ldots and addition and multiplication, and Turing is looking at a
rather strange mathematical question, which is does a program halt or
not.  It's a mathematical question {\bf that did not exist} at the
time of G\"odel's original paper.  So you see, Turing worked with
completely new concepts\ldots
 
But G\"odel's paper is not only tremendously clever, he had to have
the courage to imagine that Hilbert might be wrong.  There's another
famous mathematician of that time, von Neumann---whose grave I found
near G\"odel's, by the way, at Princeton.  Von Neumann was probably as
clever as G\"odel or anyone else, but it never occurred to him that
Hilbert could be wrong.  And the moment that he heard G\"odel explain
his result, von Neumann immediately appreciated it and immediately
started deducing consequences.  But von Neumann said, ``I missed it, I
missed the boat, I didn't get it right!''  And G\"odel did, so he was
much more profound\ldots
 
Now Turing's paper is also full of technical details, like G\"odel's
paper, because there is a programming language in Turing's paper, and
Turing also gives a rather large program, which of course has bugs,
because he wasn't able to run it and debug it---it's the program for a
universal Turing machine.  But the basic thing is the ideas, and the
new ideas in Turing's work are just breathtaking!  So I think that
Turing went beyond G\"odel, but you have to recognize that G\"odel
took the first step, and the first step is historically the most
difficult one and takes the most courage.  To imagine that Hilbert
could be wrong, which never occurred to von Neumann, that was
something!
 
\section*{I Discover Randomness in Pure Mathematics}
 
Okay, so then what happened?  Then World War II begins.  Turing starts
working on cryptography, von Neumann starts working on how to
calculate atom bomb detonations, and people forget about
incompleteness for a while.
 
This is where I show up on the scene.  The generation of
mathematicians who were concerned with these questions basically
passes from the scene with World War II.  And I'm a kid in the 1950s
in the United States reading the original article by Nagel and Newman
in {\it Scientific American\/} in 1956 that became their book.
 
And I didn't realize that mathematicians really preferred to forget
about G\"odel and go on working on their favorite problems.  I'm
fascinated by incompleteness and I want to understand it.  G\"odel's
incompleteness result fascinates me, but I can't really understand it,
I think there's something fishy\ldots As for Turing's approach, I
think it goes much deeper, but I'm still not satisfied, I want to
understand it better.
 
And I get a funny idea about randomness\ldots I was reading a lot of
discussions of another famous intellectual issue when I was a
kid---not the question of the foundations of mathematics, the question
of the foundations of {\bf physics}!  These were discussions about
relativity theory and cosmology and even more often about quantum
mechanics, about what happens in the atom.  It seems that when things
are very small the physical world behaves in a completely crazy way
that is totally unlike how objects behave here in this classroom.  In
fact things are {\bf random}---intrinsically unpredictable---in the
atom.
 
Einstein hated this.  Einstein said that ``God doesn't play dice!''
By the way, Einstein and G\"odel were friends at Princeton, and they
didn't talk very much with anybody else, and I heard someone say that
Einstein had brainwashed G\"odel against quantum mechanics!
[Laughter] It was the physicist John Wheeler, who told me that he once
asked G\"odel if there could be any connection between quantum
uncertainty and G\"odel's incompleteness theorem, but G\"odel refused
to discuss it\ldots
 
Okay, so I was reading about all of this, and I began to wonder---in
the back of my head I began to ask myself---could it be that there was
also randomness in pure mathematics?
 
The idea in quantum mechanics is that randomness is fundamental, it's
a basic part of the universe.  In normal, everyday life we know that
things are unpredictable, but in theory, in Newtonian physics and even
in Einstein's relativity theory---that's all called {\it classical\/}
as opposed to {\it quantum\/} physics---in theory in classical physics
you can predict the future.  The equations are {\it deterministic,}
not {\it probabilistic.} If you know the initial conditions exactly,
with infinite precision, you apply the equations and you can predict
with infinite precision any future time and even in the past, because
the equations work either way, in either direction.  The equations
don't care about the direction of time\ldots
 
This is that wonderful thing sometimes referred to as {\it Laplacian
determinism.} I think that it's called that because of Laplace's {\it
Essai Philosophique sur les Probabilit\'es,} a book that was published
almost two centuries ago.  At the beginning of this book Laplace
explains that by applying Newton's laws, in principle a demon could
predict the future arbitrarily far, or the past arbitrarily far, if it
knew the exact conditions at the current moment.  This is not the type
of world where you talk about free will and moral responsibility, but
if you're doing physics calculations it's a great world, because you
can calculate everything!
 
But in the 1920s with quantum mechanics it began to look like God
plays dice in the atom, because the basic equation of quantum
mechanics is the Schr\"odinger equation, and the Schr\"odinger
equation is an equation that talks about the {\bf probability} that an
electron will do something.  The basic quantity is a probability and
it's a wave equation saying how a probability wave interferes with
itself.  So it's a completely different kind of equation, because in
Newtonian physics you can calculate the precise trajectory of a
particle and know exactly how it's going to behave.  But in quantum
mechanics the fundamental equation is an equation dealing with
probabilities!  That's it, that's all there is!
 
You {\bf can't know} exactly where an electron is and what its
velocity vector is---exactly what direction and how fast it's going.
It doesn't have a specific state that's known with infinite precision
the way it is in classical physics.  If you know very accurately where
an electron is, then its velocity---its momentum---turns out to be
wildly uncertain.  And if you know exactly in which direction and at
what speed it's going, then its position becomes infinitely uncertain.
That's the infamous {\it Heisenberg uncertainty principle,} there's a
trade-off, that seems to be the way the physical universe works\ldots
 
It's an interesting historical fact that before people used to hate
this---Einstein hated it---but now people think that they can {\bf
use} it!  There's a crazy new field called {\it quantum computing\/}
where the idea is to stop fighting it.  If you can't lick them, join
them!  The idea is that maybe you can make a brand new technology
using something called {\it quantum parallelism.}  If a quantum
computer is uncertain, maybe you can have it uncertainly do many
computations at the same time!  So instead of fighting it, the idea is
to use it, which is a great idea.
 
But when I was a kid people were still arguing over this.  Even though
he had helped to create quantum mechanics, Einstein was still fighting
it, and people were saying, ``Poor guy, he's obviously past his
prime!''
 
Okay, so I began to think that maybe there's also randomness in pure
mathematics.  I began to suspect that maybe that's {\bf the real
reason for incompleteness}.  A case in point is elementary number
theory, where there are some very difficult questions.  Take a look at
the prime numbers.  Individual prime numbers behave in a very
unpredictable way, if you're interested in their detailed structure.
It's true that there are {\bf statistical} patterns.  There's a thing
called {\it the prime number theorem\/} that predicts fairly
accurately the over-all average distribution of the primes.  But as
for the detailed distribution of individual prime numbers, that looks
pretty random!
 
So I began to think about {\bf randomness}\ldots I began to think that
maybe that's what's really going on, maybe that's a deeper reason for
all this incompleteness.  So in the 1960s I, and independently some
other people, came up with some new ideas.  And I like to call this
new set of ideas {\it algorithmic information theory.}
\[
   \mbox{Algorithmic Information Theory}
\]
That name makes it sound very impressive, but the basic idea is just
to look at the size of computer programs.  You see, it's just a {\it
complexity measure,} it's just a kind of {\it computational
complexity\ldots}
 
I think that one of the first places that I heard about the idea of
computational complexity was from von Neumann.  Turing came up with
the idea of a computer as a mathematical concept---it's a perfect
computer, one that never makes mistakes, one that has as much time and
space as it needs to work---it's always finite, but the calculation
can go on as long as it has to.  After Turing comes up with this idea,
the next logical step for a mathematician is to study the time, the
work needed to do a calculation---its complexity.  And in fact I think
that around 1950 von Neumann suggested somewhere that there should be
a new field which looks at the {\bf time} complexity of computations,
and that's now a very well-developed field.  So of course if most
people are doing that, then I'm going to try something else!
 
My idea was not to look at the {\bf time}, even though from a
practical point of view time is very important.  My idea was to look
at the {\bf size} of computer programs, at the amount of information
that you have to give a computer to get it to perform a given task.
From a practical point of view, the amount of information required
isn't as interesting as the running time, because of course it's very
important for computers to do things as fast as possible\ldots But it
turns out that from a conceptual point of view, it's not that way at
all.  I believe that from a fundamental philosophical point of view,
the right question is to look at the {\bf size} of computer programs,
not at the {\bf time}.  Why?---Besides the fact that it's my idea so
obviously I'm going to be prejudiced!  The reason is because
program-size complexity connects with a lot of fundamental stuff in
physics.
 
You see, in physics there's a notion called {\it entropy,} which is
how disordered a system is.  Entropy played a particularly crucial
role in the work of the famous 19th century physicist Boltzmann,
\[
   \mbox{Ludwig Boltzmann}
\]
and it comes up in the field of statistical mechanics and in
thermodynamics.  Entropy measures how disordered, how chaotic, a
physical system is.  A crystal has low entropy, and a gas at high
temperature has high entropy.  It's the amount of chaos or disorder,
and it's a notion of randomness that physicists like.
 
And entropy is connected with some fundamental philosophical
questions---it's connected with the question of the {\it arrow of
time,} which is {\bf another} famous controversy.  When Boltzmann
invented this wonderful thing called statistical mechanics---his
theory is now considered to be one of the masterpieces of 19th century
physics, and all physics is now statistical physics---he ended up by
committing suicide, because people said that his theory was {\bf
obviously} wrong!  Why was it obviously wrong?  Because in Boltzmann's
theory entropy has got to increase and so there's an arrow of time.
But if you look at the equations of Newtonian physics, they're time
reversible.  There's no difference between predicting the future and
predicting the past.  If you know at one instant exactly how
everything is, you can go in either direction, the equations don't
care, there's no direction of time, backward is the same as forward.
 
But in everyday life and in Boltzmann statistical mechanics, there is
a difference between going backward and forward.  Glasses break, but
they don't reassemble spontaneously!  And in Boltzmann's theory
entropy has got to increase, the system has to get more and more
disordered.  But people said, ``You can't deduce that from Newtonian
physics!''  Boltzmann was pretending to.  He was looking at a gas.
The atoms of a gas bounce around like billiard balls, it's a billiard
ball model of how a gas works.  And each interaction is reversible.
If you run the movie backwards, it looks the same.  If you look at a
small portion of a gas for a small amount of time, you can't tell
whether you're seeing the movie in the right direction or the wrong
direction.
 
But Boltzmann gas theory says that there is an arrow of time---a
system will start off in an ordered state and will end up in a very
mixed up disordered state.  There's even a scary expression in German,
{\it heat death.}  People said that according to Boltzmann's theory
the universe is going to end up in a horrible ugly state of maximum
entropy or heat death!  This was the dire prediction!  So there was a
lot of controversy about his theory, and maybe that was one of the
reasons that Boltzmann killed himself.
 
And there is a connection between my ideas and Boltzmann's, because
looking at the size of computer programs is very similar to this
notion of the degree of disorder of a physical system.  A gas takes a
large program to say where all its atoms are, but a crystal doesn't
take as big a program, because of its regular structure.  Entropy and
program-size complexity are closely related\ldots
 
This idea of program-size complexity is also connected with the
philosophy of the scientific method.  You've heard of {\it Occam's
razor,} of the idea that the simplest theory is best?  Well, what's
a theory?  It's a computer program for predicting observations.  And
the idea that the simplest theory is best translates into saying that
a {\bf concise} computer program is the best theory.  What if there is
no concise theory, what if the most concise program or the best theory
for reproducing a given set of experimental data is {\bf the same
size} as the data?  Then the theory is no good, it's cooked up, and
the data is incomprehensible, it's random.  In that case the theory
isn't doing a useful job.  A theory is good to the extent that it
compresses the data into a much smaller set of theoretical
assumptions.  The greater the compression, the better!---That's the
idea\ldots
 
So this idea of program size has a lot of philosophical resonances,
and you can define randomness or maximum entropy as something that
cannot be compressed at all.  It's an object with the property that
basically the only way you can describe it to someone is to say ``this
is it'' and show it to them.  Because it has no structure or pattern,
there is no concise description, and the thing has to be understood as
``a thing in itself'', it's irreducible.
\[
   \mbox{Randomness = Incompressibility}
\]
 
The other extreme is an object that has a very regular pattern so you
can just say that it's ``a million 0s'' or ``half a million
repetitions of 01'', pairs 01, 01, 01 repeated half a million times.
These are very long objects with a very concise description.  Another
long object with a concise description is an {\it ephemeris,} I
think it's called that, it's a table giving the positions of the
planets as seen in sky, daily, for a year.  You can compress all this
astronomical information into a small FORTRAN program that uses
Newtonian physics to calculate where the planets will be seen in the
sky every night.
 
But if you look at how a roulette wheel behaves, then there is no
pattern, the series of outcomes cannot be compressed.  Because if
there were a pattern, then people could use it to win, and having a
casino wouldn't be such a good business!  The fact that casinos make
lots of money shows that there is no way to predict what a roulette
wheel will do, there is no pattern---the casinos make it their job to
ensure that!
 
So I had this new idea, which was to use program-size complexity to
define randomness.  And when you start looking at the size of computer
programs---when you begin to think about this notion of program-size
or information complexity instead of run-time complexity---then the
interesting thing that happens is that everywhere you turn you
immediately find incompleteness!  You immediately find things that
escape the power of mathematical reasoning, things that escape the
power of any computer program.  It turns out that they're everywhere!
 
It's very dramatic!  In only three steps we went from G\"odel, where
it's very surprising that there are limits to reasoning, to Turing,
where it looks much more natural, and then when you start looking at
program size, well, incompleteness, the limits of mathematics, it just
hits you in the face!  Why?!  Well, {\bf the very first question} that
you ask in my theory gets you into trouble.  What's that?  Well, in my
theory I measure the complexity of something by the size of the
smallest computer program for calculating it.  But how can I be sure
that I have the smallest computer program?
 
Let's say that I have a particular calculation, a particular output,
that I'm interested in, and that I have this nice, small computer
program that calculates it, and I think that it's the smallest
possible program, the most concise one that produces this output.
Maybe a few friends of mine and I were trying to do it, and this was
the best program that we came up with; nobody did any better.  But how
can you be {\bf sure}?  Well, the answer is that {\bf you can't be
sure}.  It turns out you {\bf can never be sure!} You can {\bf never}
be sure that a computer program is what I like to call {\it
elegant,} namely that it's the most concise one that produces the
output that it produces.  {\bf Never ever!} This escapes the power of
mathematical reasoning, amazingly enough.
 
But for any computational task, once you fix the computer programming
language, once you decide on the computer programming language, and if
you have in mind a particular output, there's got to be at least one
program that is the smallest possible.  There may be a tie, there may
be several, right?, but there's got to be at least {\bf one} that's
smaller than all the others.  But you can never be sure that you've
found it!
 
And the precise result, which is one of my favorite incompleteness
results, is that if you have $N$ bits of axioms, you can never prove
that a program is elegant---smallest possible---if the program is more
than $N$ bits long.  That's basically how it works.  So any given set
of mathematical axioms, any formal axiomatic system in Hilbert's
style, can only prove that {\bf finitely many} programs are elegant,
are the most concise possible for their output.
 
To be more precise, you get into trouble with an elegant program if
it's larger than a computerized version of the axioms---It's really
the size of the proof-checking program for your axioms.  In fact, it's
the size of the program that runs through all possible proofs
producing all possible theorems.  If you have in mind a particular
programming language, and you need a program of a certain size to
implement a formal axiomatic system, that is to say, to write the
proof-checking algorithm and to write the program that runs through
all possible proofs filtering out all the theorems, if that program is
a certain size in a language, and if you look at programs in that same
language that are larger, then you can never be sure that such a
program is elegant, you can never prove that such a program is elegant
using the axioms that are implemented in the same language by a
smaller program.  That's basically how it works.
 
So there are an infinity of elegant programs out there.  For any
computational task there's got to be at least one elegant program, and
there may be several, but you can never be sure except in a finite
number of cases.  That's my result, and I'm very proud of
it!---Another can of soda?  Thanks a lot!  My talk would be much more
interesting if this were wine or beer!  [Laughter]
 
So it turns out that you can't calculate the program-size complexity,
you can never be sure what the program-size complexity of anything is.
Because to determine the program-size complexity of something is to
know the size of the most concise program that calculates it---but
that means---it's essentially the same problem---then I would know
that this program is the most concise possible, I would know that it's
an elegant program, and you can't do that if the program is larger
than the axioms.  So if it's $N$ bits of axioms, you can never
determine the program-size complexity of anything that has more than
$N$ bits of complexity, which means almost everything, because almost
everything has more than $N$ bits of complexity.  Almost everything
has more complexity than the axioms that you're using.
 
Why do I say that?  The reason for using axioms is because they're
simple and believable.  So the sets of axioms that mathematicians
normally use are fairly concise, otherwise no one would believe in
them!  Which means that in practice there's this vast world of
mathematical truth out there, which is an infinite amount of
information, but any given set of axioms only captures a tiny finite
amount of this information!  And that's why we're in trouble, that's
my bottom line, that's my final conclusion, that's the real dilemma.
 
So in summary, I have two ways to explain why I think G\"odel
incompleteness is natural and inevitable rather than mysterious and
surprising.  The two ways are---that the idea of randomness in
physics, that some things make no sense, also happens in pure
mathematics, is one way to say it.  But a better way to say it, is
that mathematical truth is an infinite amount of information, but any
particular set of axioms just has a finite amount of information,
because there are only going to be a finite number of principles that
you've agreed on as the rules of the game.  And whenever any
statement, any mathematical assertion, involves more information than
the amount in those axioms, then it's very natural that it will escape
the ability of those axioms.
 
So you see, the way that mathematics progresses is you trivialize
everything!  The way it progresses is that you take a result that
originally required an immense effort, and you reduce it to a trivial
corollary of a more general theory!
 
Let me give an example involving Fermat's ``last theorem'', namely the
assertion that
\[
   x^n + y^n = z^n
\]
has no solutions in positive integers $x$, $y$, $z$, and $n$ with $n$
greater than 2.  Andrew Wiles's recent proof of this is hundreds of
pages long, but, probably, a century or two from now there will be a
one-page proof!  But that one-page proof will require a whole book
inventing a theory with concepts that are the natural concepts for
thinking about Fermat's last theorem.  And when you work with those
concepts it'll appear immediately obvious---Wiles's proof will be a
trivial afterthought---because you'll have imbedded it in the
appropriate theoretical context.
 
And the same thing is happening with incompleteness.
 
G\"odel's result, like any very fundamental basic result, starts off
by being very mysterious and complicated, with a long impenetrable
proof.  People said about G\"odel's original paper the same thing that
they said about Einstein's theory of relativity, which is that there
are less than five people on this entire planet who understand it.
The joke was that Eddington, astronomer royal Sir Arthur Eddington, is
at a formal dinner party---this was just after World War I---and he's
introduced as one of the three men who understands Einstein's theory.
And he says, ``Let's see, there's Einstein, and there's me, but who's
the other guy?''  I'm ruining this joke!  [Laughter]
 
So in 1931 G\"odel's proof was like that.  If you look at his original
paper, it's very complicated.  The details are programming details we
would say now---really it's a kind of complication that we all know
how to handle now---but at the time it looked very mysterious.  This
was a 1931 mathematics paper, and all of a sudden you're doing what
amounts to LISP programming, thirty years before LISP was invented!
And there weren't even any computers then!
 
But when you get to Turing, he makes G\"odel's result seem much more
natural.  And I think that my idea of program-size complexity and
information---really, algorithmic information content---makes
G\"odel's result seem more than natural, it makes it seem, I'd say,
obvious, inevitable.  But of course that's the way it works, that's
how we progress.
 
\section*{Where Do We Go from Here?!}
 
I should say, though, that if this were really true, if it were {\bf
that} simple, then that would be the end of the field of
metamathematics.  It would be a sad thing, because it would mean that
this whole subject is dead.  But I don't think that it is!
 
You know, I've been giving versions of this talk for many years.  I
make a career, a profession out of it!  It's tourism, it's the way I
get to see the world!  It's a nice way to travel!\ldots In these talks
I like to give examples of things that might escape the power of
normal mathematical reasoning.  And my favorite examples were Fermat's
last theorem, the Riemann hypothesis, and the four-color conjecture.
When I was a kid these were the three most outstanding open questions
in all of mathematics.
 
But a funny thing happened.  First the four-color conjecture was
settled by a computer proof, and recently the proof has been greatly
improved.  The latest version has more ideas and less computation, so
that's a big step forward.  And then Wiles settled Fermat's last
theorem.  There was a misstep, but now everyone's convinced that the
new proof is correct.
 
In fact, I was at a meeting in June 1993, when Wiles was presenting
his proof in Cambridge.  I wasn't there, but I was at a meeting in
France, and the word was going around by e-mail that Wiles had done
it.  It just so happened that I was session chairman, and at one point
the organizer of the whole meeting said, ``Well, there's this rumor
going around, why don't we make an announcement.  You're the session
chairman, you do it!''  So I got up and said, ``As some of you may
have heard, Andrew Wiles has just demonstrated Fermat's last
theorem.''  And there was {\bf silence!} But afterwards two people
came up and said, ``You were joking, weren't you?''  [Laughter] And I
said, ``No, I wasn't joking.''  It wasn't April 1st!
 
Fortunately the Riemann hypothesis is still open at this point, as far
as I know!
 
But I was using Fermat's last theorem as a possible example of
incompleteness, as an example of something that might be beyond the
power of the normal mathematical methods.  I needed a good example,
because people used to say to me, ``Well, this is all very well and
good, AIT is a nice theory, but give me an example of a specific
mathematical result that you think escapes the power of the usual
axioms.''  And I would say, well, maybe Fermat's last theorem!
 
So there's a problem.  Algorithmic information theory is very nice and
shows that there are lots of things that you can't prove, but what
about individual mathematical questions?  How about a natural
mathematical question?  Can these methods be applied?  Well, the
answer is no, my methods are not as general as they sound.  There are
technical limitations.  I can't analyze Fermat's last theorem with
these methods.  Fortunately!  Because if I had announced that my
methods show that Fermat's last theorem can't be settled, then it's
very embarrassing when someone settles it!
 
So now the question is, how come in spite of these negative results,
mathematicians are making so much progress?  How come mathematics
works so well in spite of incompleteness?  You know, I'm not a
pessimist, but my results have the wrong kind of feeling about them,
they're much too pessimistic!
 
So I think that a very interesting question now is to look for
positive results\ldots There are already too many negative results!
If you take them at face value, it would seem that there's no way to
do mathematics, that mathematics is impossible.  Fortunately for those
of us who do mathematics, that doesn't seem to be the case.  So I
think that now we should look for positive results\ldots The
fundamental questions, like the questions of philosophy, they're
great, because you never exhaust them.  Every generation takes a few
steps forward\ldots So I think there's a lot more interesting work to
be done in this area.
 
And here's another very interesting question: Program size is a
complexity measure, and we know that it works great in
metamathematics, but does it have anything to do with complexity in
the real world?  For example, what about the complexity of biological
organisms?  What about a theory of evolution?
 
Von Neumann talked about a general theory of the evolution of life.
He said that the first step was to define complexity.  Well, here's a
definition of complexity, but it doesn't seem to be the correct one to
use in theoretical biology.  And there is no such thing as theoretical
biology, not yet!
 
As a mathematician, I would love it if somebody would prove a general
result saying that under very general circumstances life has to
evolve.  But I don't know how you define life in a general
mathematical setting.  We know it when we see it, right?  If you crash
into something alive with your car, you know it!  But as a
mathematician I don't know how to tell the difference between a
beautiful deer running across the road and the pile of garbage that my
neighbor left out in the street!  Well, actually that garbage is
connected with life, it's the debris produced by life\ldots
 
So let's compare a deer with a rock instead.  Well, the rock is
harder, but that doesn't seem to go to the essential difference that
the deer is alive and the rock is a pretty passive object.  It's
certainly very easy for us to tell the difference in practice, but
what is the fundamental difference?  Can one grasp that
mathematically?
 
So what von Neumann was asking for was a general mathematical theory.
Von Neumann used to like to invent new mathematical theories.  He'd
invent one before breakfast every day: the theory of games, the theory
of self-reproducing automata, the Hilbert space formulation of quantum
mechanics\ldots Von Neumann wrote a book on quantum mechanics using
Hilbert spaces---that was done by von Neumann, who had studied under
Hilbert, and who said that this was the right mathematical framework
for doing quantum mechanics.
 
Von Neumann was always inventing new fields of mathematics, and since
he was a childhood hero of mine, and since he talked about G\"odel and
Turing, well, I said to myself, if von Neumann could do it, I think
I'll give it a try.  Von Neumann even suggested that there should be a
theory of the complexity of computations.  He never took any steps in
that direction, but I think that you can find someplace where he said
that this has got to be an interesting new area to develop, and he was
certainly right.
 
Von Neumann also said that we ought to have a general mathematical
theory of the evolution of life\ldots But we want it to be a very
general theory, we don't want to get involved in low-level questions
like biochemistry or geology\ldots He insisted that we should do
things in a more general way, because von Neumann believed, and I
guess I do too, that if Darwin is right, then it's probably a very
general thing.
 
For example, there is the idea of {\it genetic programming,} that's
a computer version of this.  Instead of writing a program to do
something, you sort of evolve it by trial and error.  And it seems to
work remarkably well, but can you prove that this has got to be the
case?  Or take a look at Tom Ray's {\it Tierra\ldots} Some of these
computer models of biology almost seem to work too well---the problem
is that there's no theoretical understanding why they work so well.
If you run Ray's model on the computer you get these parasites and
hyperparasites, you get a whole ecology.  That's just terrific, but as
a pure mathematician I'm looking for theoretical understanding, I'm
looking for a general theory that starts by defining what an organism
is and how you measure its complexity, and that {\bf proves} that
organisms have to evolve and increase in complexity.  That's what I
want, wouldn't that be nice?
 
And if you could do that, it might shed some light on how general the
phenomenon of evolution is, and whether there's likely to be life
elsewhere in the universe.  Of course, even if mathematicians never
come up with such a theory, we'll probably find out by visiting other
places and seeing if there's life there\ldots But anyway, von Neumann
had proposed this as an interesting question, and at one point in my
deluded youth I thought that maybe program-size complexity had
something to do with evolution\ldots But I don't think so anymore,
because I was never able to get anywhere with this idea\ldots
 
So I think that there's a lot of interesting work to be done!  And I
think that we live in exciting times.  In fact, sometimes I think that
maybe they're even a little bit too exciting!\ldots And I hope that if
this talk were being given a century from now, in 2099, there would be
{\bf another} century of exciting controversy about the foundations of
mathematics to summarize, one with different concerns and
preoccupations\ldots It would be interesting to hear what that talk
would be like a hundred years from now!  Maybe some of you will be
there!  Or give the talk even!  Thank you very much!  [Laughter \&
Applause]
 
\section*{Further Reading}
 
\begin{enumerate}
 
\item
G. J. Chaitin, {\it The Unknowable,}
Springer-Verlag, 1999.
 
\item
G. J. Chaitin, {\it The Limits of Mathematics,}
Springer-Verlag, 1998.
 
\end{enumerate}
 
\end{document}